\documentclass[doublecol]{epl2}
\usepackage{graphicx}
\usepackage{amssymb,amsmath}
\usepackage{subfigure}
\usepackage{hyperref}

\title{Power spectra of a constrained totally asymmetric simple exclusion process}
\shorttitle{Power spectra of a constrained TASEP}
\author{L. Jonathan Cook \and R. K. P. Zia}
\shortauthor{L. Jonathan Cook \etal}
\institute{Department of Physics, Virginia Tech, Blacksburg, VA 24061, USA}

\abstract{
To synthesize proteins in a cell, an mRNA has to work with a finite pool of
ribosomes. When this constraint is included in the modeling by a totally asymmetric
simple exclusion process (TASEP), non-trivial consequences emerge. Here, we
consider its effects on the power spectrum of the total occupancy, through Monte
Carlo simulations and analytical methods. New features, such as dramatic
suppressions at low frequencies, are discovered. We formulate a theory based
on a linearized Langevin equation with discrete space and time. The good
agreement between its predictions and simulation results provides some
insight into the effects of finite resoures on a TASEP.}

\begin{document}
\maketitle

%\email{lacook1@vt.edu}

%\email{rkpzia@vt.edu}

%\email{schmittm@vt.edu}

\section{Introduction}
Unlike systems in thermal equilibrium, those driven far
from equilibrium are poorly understood. Even when they settle into time
independent, steady states, there is no comprehensive framework that
provides, say, the stationary distributions. Yet, such systems are
ubiquitous in nature, from small biological systems like the cell to global
networks like the internet. In order to gain some insight into the physics
of such systems, we may first turn to simple models, from which more complex
theories may be built. The totally asymmetric simple exclusion process
(TASEP) \cite{TASEP} is a prime example, playing a role similar to the Ising
model for the understanding of phase transitions in equilibrium statistical
mechanics. Since its stationary distributions are known analytically, many interesting macroscopic properties can be computed exactly \cite{Exact-sol}. Beyond the simplest TASEP, many generalizations have been
proposed for modeling a variety of systems in science and engineering, from
protein synthesis \cite{Pro} and surface growth \cite{Sur} to vehicular
traffic \cite{Traf}.

Despite these advances, many of its behaviors continue to present us with
surprises. One recent example is the power spectrum associated with the total particle occupancy showing dramatic oscillations (in the frequency domain) \cite{ASZ07}. Although the predictions from a linearized Langevin equation are
found to agree well with simulation data, puzzles concerning some of the fit
parameters linger \cite{ASZ07}. In this study, we pursue this quantity
further, but in a different setting. In modeling of protein synthesis, a
TASEP represents a mRNA while the particles represent ribosomes that bind to
one end of the mRNA, move to the other end, and detach. In solvable
TASEPs, particles enter the system with a fixed rate, as if coupled to an
infinite reservoir. Yet, in a real cell, the number of ribosomes is clearly
finite. Thus, the effects of feedback from ``ribosome recycling'' \cite{TC-recycling} deserve attention, especially if the effective binding rate
may depend on the concentration of the available ribosomes. Recently, two
studies investigated the effects of a \emph{finite} pool of particles on
TASEPs \cite{ASZ08,CZ09}, given reasonable assumptions on how the effective
entry rate depends on the numbers in the pool. Here, we address another
issue: How are the power spectra affected? Using Monte Carlo simulations, we
found serious suppressions at low frequencies. In the remainder of this
letter, we briefly discuss the model, the simulation results, and a
theoretical explanation for this suppression. We end with an outlook for
further research.

\section{Model}
The original open TASEP, which will be referred to here as
``ordinary,'' consists of a one-dimensional lattice with sites labeled by $%
x\in \left[ 1,L\right] $. Each site may be occupied by at most one particle.
In each time step, a randomly chosen particle hops to the next site ($%
x\rightarrow x+1$), if the latter is empty. If chosen, the last particle
leaves the lattice with probability $\beta $, while the first site, if
empty, will be filled by a particle with probability $\alpha $. Thus, the
total occupancy is a fluctuating quantity in time: $N\left( t\right) $. In
the steady state, the time average ($\bar{N}$) and overall density ($\bar{\rho}\equiv \bar{N}/L$) settle to constants. In the thermodynamic limit,
three phases exist: (i) maximal current (MC) with $\bar{\rho}=1/2$ for $\alpha,\beta \geq 1/2$, (ii) high density (HD) with $\bar{\rho}=1-\beta$
for $\beta <1/2$ and $\alpha >\beta $, and (iii) low density (LD) with $\bar{%
\rho}=\alpha $ for $\alpha <1/2$ and $\alpha <\beta $. At the HD-LD phase
boundary ($\alpha =\beta <1/2$), a LD region (for smaller $x$) co-exists
with a HD region, separated by an interface with microscopic width, known as
the ``shock'' or the domain wall (DW). As the DW performs a random walk
throughout the lattice, the overall density suffers macroscopic
fluctuations, but $\bar{\rho}$ is restored to $1/2$ over long times. Though
this HD-LD boundary is not a phase, it is often referred to as the ``shock
phase'': (SP).

In the constrained TASEP, we connect the lattice to a ``pool'' of $N_p$
particles. When the last particle leaves the lattice (with rate $\beta $),
it is recycled back into this pool, while a
particle enters the lattice (if the first site is empty) with a $N_p$%
-dependent rate $\alpha _{eff}$. Thus, the total number of particles in the
system, $N_{tot}=N+N_p$, is a constant. Following previous studies \cite{ASZ08,CZ09}, we use 
\begin{equation}
\alpha _{eff}\left( N_p\right) =\alpha \tanh \left( N_p/N^{*}\right)
\label{alpha-eff}
\end{equation}
where $N^{*}$ marks a cross-over scale, chosen typically as $\bar{\rho}L$.
This form is motivated by the following. If there are few ribosomes
available, the binding rate should be proportional to their concentration.
Thus, $\alpha _{eff}\propto N_p$ for small $N_p$. On the other hand,
regardless of how abundant the ribosomes are, the binding rate should not
exceed some intrinsic value, which we modeled by $\alpha $. For convenience,
we will label the pool as a site ($x=0$) on a \emph{periodic} lattice with $%
L+1$ sites. Clearly, there is no exclusion for this site (unrestricted $N_p$%
) and the rules of hopping into/out of it differ from the rest.

For our simulation study, time $t$ is measured in Monte Carlo Steps (MCS).
In one MCS, $L+1$ nearest-neighbor pairs of sites are chosen randomly for an
update attempt. Starting with all particles being in the pool, the system is
allowed to reach steady state by waiting 100k MCS typically. Thereafter, the
total number of particles on the lattice, $N\left( t\right) $, is measured
every $\ell $ MCS until $T$ MCS. Using mostly $\ell ,T=10^2,10^6$, we
compute the Fourier transform $\tilde{N}(m)=\sum_{\tau =1}^{T/\ell }N\left(
\ell \tau \right) e^{2\pi im\tau \ell /T}$ with $m\in \left[ 1,T/\ell
\right] .$ Typically 100 such runs are carried out and our power spectrum is
defined as $I(m)=\langle |\tilde{N}(m)|^2\rangle$, 
%\begin{equation}
%I(m)=\langle |\tilde{N}(m)|^2\rangle
%\end{equation}
where $\langle \dots \rangle $ indicates the average over these runs.

\section{Simulation results}
With appropriate $\alpha $ and $\beta $, the
ordinary TASEP settles into a steady state: MC, LD, HD, or SP. For the
constrained TASEP, we have an additional control parameter: $N_{tot}$. In 
\cite{ASZ08}, the effects of tuning $N_{tot}$ on $\bar{\rho}$ (and the
overall current) were reported. Not surprisingly, the TASEP is essentially
in a LD phase when $N_{tot}$ is small, regardless of ($\alpha $,$\beta $).
As $N_{tot}$ is increased, more interesting crossover behaviors were found \cite{ASZ08}. Here, we turn to $I(m)$, which provides information on
time correlations $\langle N\left( t\right) N\left( t^{\prime }\right)
\rangle $. To distinguish the two power spectra, we will use $I_o$ for the
ordinary TASEP \cite{ASZ07} and $I_c$ for the constrained case here. In all
cases, $I_c$ is severely smaller than $I_o$ at low frequencies. We will
focus on two regimes: (i) the LD and (ii) crossing over to the HD, through
the SP.

\begin{figure}[tb]
\begin{center}
\includegraphics[width=8.6cm]{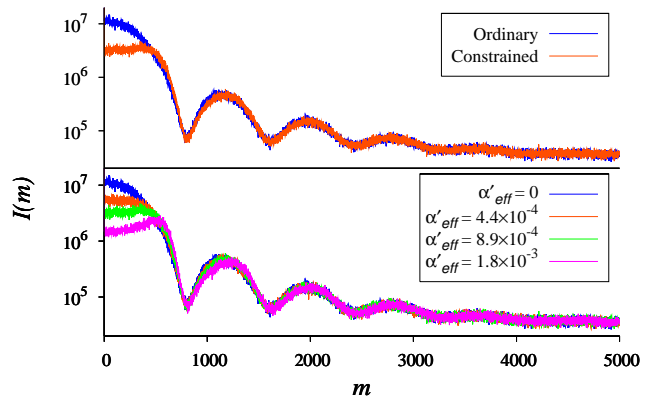}
\end{center}
%\vspace*{-0.75cm}
\caption{(Color online) Upper: Comparison of the power spectra of the ordinary and constrained TASEP with $\rho=0.1$ and $L=1000$. Lower: Constrained power spectra for $\rho =0.1$ with $L=1000$ with
different $\alpha^{\prime}_{eff}$. The $\alpha^{\prime}
_{eff}=0$ line corresponds to the ordinary TASEP with a
constant $\alpha$.}
%\vspace*{-0.5cm}
\label{vary-f-prime}
\end{figure}

For the LD regime, $I_o(m)$ displays dramatic oscillations which depend on $L$
and $\alpha $. Similar oscillations are found in $I_c(m)$, except for the
suppression at low $m$'s. An example is shown in fig.\ \ref{vary-f-prime}, where $%
\beta =0.9$ and $L=1000$ in both cases, while $\alpha =0.1$ for $I_o(m)$ and 
$\left( \alpha ,N_{tot},N^{*}\right) =\left( 0.3,205,300\right) $ for $%
I_c(m) $. For large $m$'s, the two spectra are approximately equal, since we
have chosen the $\alpha $ for $I_o$ to be approximately the average $\alpha _{eff}$
for $I_c$. This behavior can be intuitively understood: On short time
scales, fluctuations in $N\left( t\right) $ do not have time to traverse
the entire lattice and to contribute to the feedback, via $\alpha
_{eff}\left( N_{tot}-N\right) $. By contrast, fluctuations over longer times
will affect $\alpha _{eff}$ and so, $I_c(m)$ at low $m$. Further, we can
expect a significant role from the ``stiffness'' of $\alpha _{eff}$, i.e., $%
\alpha _{eff}^{\prime }\equiv \left. \partial _{N_p}\alpha _{eff}\right| _{%
\bar{N}_p}$ , where $\bar{N}_p=$ $N_{tot}-\bar{\rho}L$ is the average pool
occupation. To explore further how $I_c$ is affected by a range of $\alpha
_{eff}^{\prime }$, we investigated eq.\ (\ref{alpha-eff}) with different $%
N^{*}$'s (and other parameters so as to keep $\alpha _{eff}$ close to 0.1).
The results are illustrated in fig.\ \ref{vary-f-prime} and confirm our
belief that a larger $\alpha _{eff}^{\prime }$ induces a larger suppression. 
%\begin{figure}[tbh]
%\begin{center}
%\includegraphics[width=8.6cm]{L1000-LD.eps}
%\end{center}
%\caption{Constrained and ordinary power spectra for $\rho =0.1$ with $L=1000$
%plotted as a function of $m=\omega T/2\pi $.}
%\label{L1000-LD}
%\end{figure}

For the other regime -- crossover through SP to HD, $I_c$ is also severely
suppressed at low $m$, though the detailed structures are less interesting.
We should emphasize that, in this SP-like regime, the average $\alpha _{eff}$
is equal to $\beta $, for a range of $N_{tot}$ values. Thus, we may expect $%
I_c\sim m^{-2}$, just like in $I_o$ \cite{ASZ07}. Similar to above, this
expectation is indeed valid, but only for large $m$.  In the suppressed region at low $m$, $I_c$ approaches a constant value.
%The suppression at low $m$ is illustrated in Fig.\ \ref{power-compare-SP}, where an appropriate $I_o$ is also shown for comparison. 
%\begin{figure}[tb]
%\begin{center}
%\includegraphics[width=8.6cm, height=5cm]{power-compare-SP.eps}
%\end{center}
%\vspace*{-0.75cm}
%\caption{(Color online) Constrained and ordinary power spectra for $L=1000$ when the
%average $\alpha _{eff}=\beta$.}
%\vspace*{-0.5cm}
%\label{power-compare-SP}
%\end{figure}

\section{Theoretical considerations}
Since $I\left( m\right) $ carries
information on time correlations of $N\left( t\right) $, even $I_o$ cannot
be accessed from the exact stationary distribution \cite{Exact-sol}. Indeed,
finding $I_o$ would be still a serious challenge \cite{Blythe}, even if all
the eigenvalues and the (left and right) eigenvectors of the Liouvillian (of
the master equation governing TASEP) were explicitly known \cite{Exact-dyn}!
Thus, we will attempt to understand the suppression through a
semi-phenomenological approach, similar to the one in \cite{ASZ07}. Also
based on a Langevin equation for the local particle density $\rho (x,t)$,
our approach here is improved over \cite{ASZ07} in several ways. One is the use of discrete
space and time points \cite{PPOF}, thus avoiding any issues associated with
UV cutoffs. Another concerns the use of a ring of $L+1$ sites rather than
open boundaries, so that any issues associated with boundary currents and
noise are avoided. Finally, due to the constraint, $N\left( t\right)
=N_{tot}-N_p\left( t\right) $, so that $I\left( m\right) $ is directly
related to the fluctuations of $\rho (0,t)$, which is the pool.

For simplicity, we define $\Delta _t\rho (x,t)\equiv \rho (x,t+1)-\rho (x,t)$
and suppress $t$ when no ambiguity exists. For all but the ``exceptional
sites,'' i.e., $x=0,1,$ and $L$, we may write 
\begin{eqnarray}
\Delta _t\rho (x) &=&\rho (x-1)[1-\rho (x)]-\rho (x)[1-\rho (x+1)]  \nonumber
\\
&&+\xi (x-1,t)-\xi (x,t)  \label{LE}
\end{eqnarray}
where $\xi (x,t)$ is the noise associated with the jump $x\rightarrow $ $x+1$%
. Next, we define $\tilde{\xi}\equiv \xi (L)-\xi (0)$ and write 
\begin{equation}
\Delta _t\rho (0)=\beta \rho (L)-\alpha _{eff}\left( \rho (0)\right) [1-\rho
(1)]+\tilde{\xi}\,  \label{LE-Np}
\end{equation}
and similar equations for $\rho (1)$ and $\rho (L)$. The conservation
constraint, $\sum_x\Delta _t\rho (x)=0$, can be verified. In principle, it
is possible for $\rho (x\neq 0)$ to exceed unity (due to the Gaussian noise
and discrete time), but we will disregard this complication here.

To simplify the computation, we choose $\alpha $, $\beta $, and $N_{tot}$
such that the system is in a LD state with a \emph{flat} (average)\emph{\
profile}, $\bar{\rho}$. For fluctuations in the lattice ($x\in \left[
1,L\right] $), we write $\varphi (x,t)=\rho (x,t)-\bar{\rho}$ and, for the
pool, $\zeta (t)=N_p(t)-\bar{N}_p$ (due to the conserved dynamics, $\zeta +\sum_x\varphi \equiv 0$). Next, we expand the equations above, keeping only
linear terms in $\varphi $ and $\zeta $. Of course, we must also linearize $%
\alpha _{eff}$ around its average (which is, in LD, just $\bar{\rho}$): 
\begin{equation}
\alpha _{eff}=\bar{\rho}-\alpha _{eff}^{\prime }\zeta (t)+\dots  \label{zeta}
\end{equation}
The result is a standard biased diffusion equation for $\varphi $ within the
lattice, along with novel equations for the exceptional sites. In this
approach, the diffusion coefficient is just $1/2$ and the bias is $v\equiv
1-2\bar{\rho}$. Note that the only difference between $I_o
$ and $I_c$ lies in $\alpha _{eff}^{\prime }$ in eq.\ (\ref{zeta}) being
zero or not. This set of linearized equations can be solved in Fourier
space, with $\tilde{\varphi}(k,\omega )$, $\tilde{\zeta}(\omega )$, and $%
\tilde{\xi}(k,\omega )$. Deferring details to elsewhere, we present only the
results along with a few remarks here.

Assuming Gaussian noise with zero mean and correlation $\langle \tilde{\xi}%
(k,\omega )\tilde{\xi}(k^{^{\prime }},\omega ^{^{\prime }})\rangle =A\delta
_{k,k^{^{\prime }}}\delta _{\omega ,\omega ^{^{\prime }}}$, we solve the
linearized equations and obtain, in particular,
\begin{equation}
\langle |\tilde{\zeta}(\omega )|^2\rangle =\sum_k\frac{2A(1-\cos k)}{%
|Q(\omega )P(k,\omega )|^2}  \label{zeta2}
\end{equation}
where $P(k,\omega )=e^{i\omega }-\cos k+iv\sin k$ and 
\begin{equation}
Q(\omega )=\sum_q\frac{e^{i\omega }-1+\alpha _{eff}^{\prime }\left( 1-\bar{%
\rho}\right) (1-e^{-iq})}{(L+1)P(q,\omega )}  \label{Q}
\end{equation}
At this level of approximation, $A$ is usually taken as $\bar{\rho}\left( 1-%
\bar{\rho}\right) $\cite{A}. Recalling that the complete power spectrum is
precisely $\langle |\tilde{\zeta}(\omega )|^2\rangle $, we can compare (\ref
{zeta2}) with the simulation data. However, first we must account for a minor complication:
the measurements taken only every $\ell $ MCS. Summing over the unobserved $t
$'s is equivalent to summing over the harmonics $\omega _{m,\mu }=2\pi
\left( \frac mT+\frac \mu \ell \right) ;$ $\mu =0,...,\ell -1$. A
straightforward computation leads to 
\begin{equation}
I(m)=\frac 1{\ell ^2}\sum_{\mu =0}^{\ell -1}\sum_k\frac{2A(1-\cos k)}{%
|Q(\omega _{m,\mu })P(k,\omega _{m,\mu })|^2}.  \label{PS-theory}
\end{equation}
Note that the only difference between $I_o$ and $I_c$ is the extra term
proportional to $\alpha _{eff}^{\prime }$ in $Q(\omega )$.

\begin{figure}[tb]
\begin{center}
\includegraphics[width=8.6cm, height=5cm]{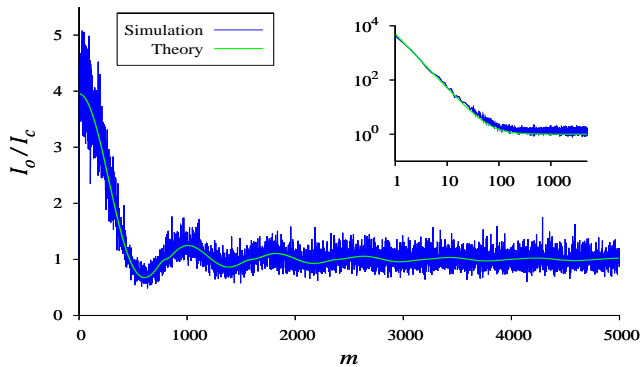}
\end{center}
%\vspace*{-0.75cm}
\caption{(Color online) Simulation data and theory for the ratio of $I_o$ and $I_c$ plotted as a function of $m$ for LD (inset for crossover in HD).}
%\vspace*{-0.5cm}
\label{power-ratio}
\end{figure}

In previous studies \cite{ASZ07}, a similar approach led to good fits, \emph{%
provided }an effective ``renormalized'' diffusion coefficient $D$ is introduced and allowed as a
fit parameter.  Our result, as shown in eq.\ (\ref{PS-theory}),
appears more complicated than the previous approach. But, when it is
evaluated numerically, it is very similar to the earlier prediction \cite{ASZ07} for $I_o\left( m\right) $. However, our UV cutoff is built in, so
that there is no obvious way to insert a ``renormalized'' $D$ into either
eq.\ (\ref{LE}) or $P(k,\omega )$. To go beyond this impasse, we plan to
pursue perturbation theory seriously. Despite these unresolved issues, let
us present a remarkable prediction from eq.\ (\ref{PS-theory}) here. Instead
of the individual $I$'s, consider the ratio $I_0/I_c$. For reasons yet to be
understood, there is quite good agreement between the data and theoretical
results. In fig.\ \ref{power-ratio}, we illustrate this ``zero parameter
fit'' for these ratios in a case with $L=1000$. We should caution that the
agreement in a larger system ($L=32000$), though adequate, is not as
impressive. Thus, the best we may conclude at this stage is that the ratio $%
I_0/I_c$ is somehow less sensitive to renormalization effects. 

Lastly, we turn to the crossover regime in the HD case, where a DW wanders
relatively freely within the lattice. The appropriate comparison case in the
ordinary TASEP is SP where $I_o\sim m^{-2}$ \cite{ASZ07}. Our goal here is to
explain the suppression of $I_c$ at small $m$. A physically intuitive
picture is that the DW is localized by the feedback (through $\alpha _{eff}$%
) in the constrained TASEP \cite{CZ09}. Thus, its dynamics must be well
described by a noisy harmonic oscillator and $I_c$ should be just a
Lorentzian. For a quantitative picture, we start with eq.\ (\ref{LE-Np}) for 
$N_p\left( t\right) =\rho (0,t)$, but with the simplifying approximations $%
\rho (L)\cong 1-\beta $ and $\rho (1)\cong \alpha _{eff}$.
%\begin{equation}
%\Delta _tN_p=\beta (1-\beta )-\alpha _{eff}\left( N_p\right) [1-\alpha
%_{eff}\left( N_p\right) ]-\tilde{\xi}\text{.}  \label{LE-SP}
%\end{equation}
For the SP in an ordinary TASEP, $\alpha _{eff}\equiv \alpha =\beta $, so
that $\Delta _tN_p=\xi$ (with appropriate boundary conditions),
leading to $I_o\propto m^{-2}$. With finite resources, a range of $N_{tot}$
produces $\alpha _{eff}\left(\bar{N}_p\right)=\beta $, and so, we expand
the second term to $O\left( \zeta \right) $. In the continuous $t$ limit,
eq.\ (\ref{LE}) becomes $\partial _t\zeta =-\gamma \zeta +\xi$ with $%
\gamma =(1-2\beta )\alpha _{eff}^{\prime }\left( \bar{N}_p\right) $. The
solution is trivial, leading to $\langle |\tilde{\zeta}(\omega )|^2\rangle
\propto \left( \omega ^2+\gamma ^2\right) ^{-1}$. As above, we compare the
ratios $I_o/I_c$ , for which the theory predicts 
\begin{equation}
\frac{I_o(m)}{I_c(m)}\propto 1+\left( \frac{\gamma T}{2\pi m}\right) ^2\text{%
.}  \label{ratio}
\end{equation}
In fig.\ \ref{power-ratio}, we plot this ratio from simulation data (for $(L,\alpha,\beta,N^*,N_{tot})$=$(1000,0.75,0.25,750,800)$)
as well as the expression above. The agreement is remarkably good,
especially since \emph{no fit parameters} (apart from the overall proportionality constant) have been introduced. We have made such
comparisons for other values of $\alpha ,\beta ,$ etc. and all are similar,
giving us confidence that this theory is quite adequate for predicting how
the feedback mechanism suppresses $I_c(m)$ at small $m$.

%\begin{figure}[tb]
%\begin{center}
%\includegraphics[width=8.6cm]{power-theory-SP.eps}
%\end{center}
%\caption{Simulation data and theory for the constrained TASEP power spectrum
%for $\alpha =0.7$, $\beta =0.3$ and $N_{tot}=800$ with $L=1000$ plotted as a
%function of $m$.}
%\label{power-theory-SP}
%\end{figure}

\section{Summary and outlook}
We studied the power spectrum associated with
the total occupation, $N\left( t\right) $, on a TASEP constrained by finite
resources, using both simulations and analytic techniques. Considerable
suppression at low frequencies is found, depending on the feedback through $%
\alpha _{eff}$. Though much improved over a previous approach, the theory we
studied also predicts only certain aspects of the observed spectra.
Nevertheless, it is surprising that there is good agreement for the \emph{%
ratio} of the power spectra ($I_o/I_c$), with no fit parameters!

Obviously, this approach to understanding the simulation results leaves room
for improvement. We believe that the non-linear terms neglected here will be
the key to better predictions, especially at higher densities where the
excluded volume constraint should be more relevant. Hopefully, a full
investigation of their effects will also reveal the secrets of the sensitive 
$L$-dependent of $D$ and $A$ found earlier \cite{ASZ07,AZ09}. Beyond the
study of a single TASEP coupled to a finite pool of particles, we look
toward generalizations of the model which may have further implications for
protein synthesis in a real cell. In particular, we should study TASEPs with
large particles and  inhomogeneous hopping rates \cite{Pro}. We are aware of
preliminary studies of the power spectra of these generalized TASEPs \cite{JJ}
and plan systematic investigations. Further, we should consider
multiple TASEPs competing for the \emph{same} pool of particles, just as
many mRNA's in a cell compete for the same pool of ribosomes. A natural
question -- are there winners or losers? -- can be crudely answered by
studying their occupations, $N_i\left( t\right) $, and currents, $J_i\left(
t\right) $. Obviously, power spectra of various combinations can reveal
fluctuations and cross correlations between the different quantities.
Perhaps they can be exploited as a sensitive diagnostic tool for
distinguishing different mechanisms that control translation. Beyond
translation, we believe the study of power spectra will facilitate our
understanding of non-equilibrium systems in other context, such as social
networks, traffic flow, and finance.

\acknowledgments
We thank D.A. Adams, R. Blythe, J.J. Dong, B.
Schmittmann, and S. Mukherjee for illuminating discussions. This research
is supported in part by a grant from the US National Science Foundation,
DMR-0705152.
%\vspace*{-0.5cm}

\end{document}